\documentclass[12pt]{article}

\usepackage[margin = 2cm]{geometry}

\usepackage{graphicx}
\usepackage{amssymb}

\usepackage{lineno}
\usepackage[title]{appendix}

\usepackage{multirow}
\usepackage{setspace}
\usepackage{natbib}
\usepackage{booktabs}
\usepackage{caption}
\usepackage{amsmath}
\usepackage{hyperref}
\usepackage{ccicons}
\usepackage[T1]{fontenc}

\usepackage{enumitem}
\usepackage[table,xcdraw]{xcolor}
\usepackage{floatrow}
\usepackage{subcaption}
\usepackage{authblk}
\usepackage{booktabs}
\usepackage[normalem]{ulem}
\useunder{\uline}{\ul}{}
\graphicspath{ {././} }

\begin{document}


\title{Large-scale Time-Varying Portfolio Optimisation using Graph Attention Networks \footnote{\scriptsize NOTICE: This is a preprint of a published work. Changes resulting from the publishing process, such as editing, corrections,structural formatting, and other quality control mechanisms may not be reflected in this version of the document. Please cite this work as follows: Korangi, K., Mues, C. and Bravo, C. (2022). Large-scale Time-Varying Portfolio Optimisation using Graph Attention Networks. 
This work is made available under a \href{https://creativecommons.org/licenses/by/4.0/}{Creative Commons BY license}. \ccby}}

\author[1,2]{Kamesh Korangi}
\author[1,2]{Christophe Mues}
\author[3]{Cristi\'{a}n Bravo}

\affil[1]{Department of Decision Analytics and Risk, Southampton Business School, University of Southampton, University Road, SO17 1BJ, United Kingdom.}

\affil[2]{Centre for Operational Research, Management Sciences and Information Systems (CORMSIS), University of Southampton, University Road, SO17 1BJ, United Kingdom.}

\affil[3]{Department of Statistical and Actuarial Sciences, The University of Western Ontario,1151 Richmond Street, London, Ontario, N6A 5B7, Canada.}

\date{}

\maketitle
\begin{abstract}
Apart from assessing individual asset performance, investors in financial markets also need to consider how a set of firms performs collectively as a portfolio. Whereas traditional Markowitz-based mean-variance portfolios are widespread, network-based optimisation techniques have built upon these developments. However, most studies do not contain firms at risk of default and remove any firms that drop off indices over a certain time. This is the first study to incorporate risky firms and use all the firms in portfolio optimisation. We propose and empirically test a novel method that leverages Graph Attention networks (GATs), a subclass of Graph Neural Networks (GNNs). GNNs, as deep learning-based models, can exploit network data to uncover nonlinear relationships. Their ability to handle high-dimensional features and accommodate customised layers for specific purposes makes them particularly appealing for large-scale problems such as mid- and small-cap portfolio optimization. This study utilises 30 years of data on mid-cap firms, creating graphs of firms using distance correlation and the Triangulated Maximally Filtered Graph approach. These graphs are the inputs to a GAT model that we train using custom layers which impose weight and allocation constraints and a loss function derived from the Sharpe ratio, thus directly maximising portfolio risk-adjusted returns. This new model is benchmarked against a network characteristic-based portfolio, a mean variance-based portfolio, and an equal-weighted portfolio. The results show that the portfolio produced by the GAT-based model outperforms all benchmarks and is consistently superior to other strategies over a long period while also being informative of market dynamics.
\end{abstract}


\section{Introduction}

Portfolio optimisation is crucial in financial risk management, as performance correlations between firms in a portfolio bring unforeseen risks. Given that each investor's risk profile differs, the portfolio construction model must also account for different objectives. One such type of model, based on the classic work by \citet{Markowitz1952, Markowitz1959}, is mean-variance optimisation, which trades off maximising returns against minimising volatility. 
In this paper, too, we look to optimise with a mean-variance objective, but we do so over a large set of firms that could also default or go bankrupt. This problem must be solved for an investable universe that is expanding, as financial markets continue to develop, and emerging and private markets are becoming increasingly accessible. All these assets have different risk and liquidity profiles. Against this backdrop, which firms to select and what proportion of capital to allocate to each is an increasingly high-dimensional problem. 
  
Portfolio optimisation often involves estimating the expected returns and covariance matrix and then using a constrained optimisation method to find the asset allocation weights that maximise the portfolio objective. The classical mean-variance measure is not without its problems, though, and portfolios optimised using it have been shown to exhibit poor out-of-sample performance \citep{siegel2007}. Assumptions about the normality of returns and absence of transaction costs, as well as the presence of regimes in markets, make the classical model difficult to implement \citep{Guidolin2011}. Even if these assumptions are fulfilled, the expected mean of the portfolio returns and the covariance matrix cannot be readily estimated as they are not observed in practice, which means that, instead, the sample mean and covariance matrix are commonly used \citep{Ao2019}. Furthermore, it is challenging for such models to cope with high dimensionality, a common characteristic of modern portfolios \citep{DeMiguel2009a}. To better address these challenges, new methods to solve the portfolio optimisation problem continue to be developed, borrowing from different techniques in other domains, such as fuzzy programming \citep{ArenasParra2001}, cluster analysis \citep{puerto2020}, quantum annealing \citep{Venturelli2019} and deep reinforcement learning \citep{shi2022}.

Of particular interest to our work are topological or network studies for portfolio optimisation \citep{pozzi2013a,Li2019}. Network models exploit graph data structures to identify relationships that may be impossible to detect by Euclidean data-based models. In our case, the network nodes are the firms, and each edge represents a relationship between two firms. More formally, the network at a given time $t$ is represented as an undirected graph $\mathcal{G}= (\mathcal{V}_{t},\mathcal{E}_{t})$ where  
$\mathcal{V}_{t}$ are the nodes or firms and $\mathcal{E}_{t}$ is the set of edges, often represented by an adjacency matrix $A$ of dimension $|\mathcal{V}_{t}| \times |\mathcal{V}_{t}| $.
 
The aforementioned network studies consistently find that allocating capital to firms in the peripheries of the networks produces higher returns, due to low correlations with other parts of the network. Similarly, to produce a diversified portfolio, mean-variance models also tend to prefer firms in the peripheries of the network \citep{Onnela2003}. However, studies using the former methods have only been deployed to small portfolios or were limited to a specific sector. Here, we look to extend these topological analyses to the whole market of US mid-cap companies, a much more challenging and realistic problem setting.
 
Mid-cap firms (in short \lq{mid-caps}\rq) are defined as firms with a 1 to 10 billion USD market capitalisation and are likely constituents of the Dow Jones Wilshire Mid-cap or S\&P 400 Mid-cap indexes. Modelling the performance of these firms is complicated by the low-volume and, at times, illiquid nature of trading, which makes their return distributions non-normal \citep{Castellano2014}. They are also far more numerous compared to large-cap firms, which makes mid-caps less suitable for analysts to cover. However, as they behave as a separate autonomous asset class, they can further improve the diversification aspect of a portfolio if included \citep{Petrella2005}. Over the long term, mid-caps also provide a premium in return for the same risk, which is desirable for any portfolio seeking financial returns \citep{Ge2018}. Given the large number of companies in the mid-cap universe, simple index replication strategies, such as those implementing the popular market-weighted methodology of the Russell 2000 Index, can be costly though. Furthermore, investors may have different horizons and risk tolerance, whilst constituent churn can negatively affect performance \citep{Cai2008,Cremers2020}. Hence, a comprehensive approach is needed to generate portfolio weights for mid-cap firms that yield better risk-adjusted returns. Another area of practical interest to which our study may be applicable is the development of new automated ETF strategies for such companies. The latter would require large-scale portfolio optimisation models incorporating the strategy constraints of the particular ETF.
 
Studying the correlation between firm's returns or volatilities is an integral part of any portfolio optimisation procedure.  Pearson correlation approaches, generally used by mean-variance models, can only capture linear dependencies and pairwise correlations. Instead, in this work, we employ the distance correlation measure \citep{Szekely2007}. This is able to capture non-linear relationships between pairs of firms. \citet{Sun2019} compared the use of distance correlation with Pearson correlation for portfolio optimisation, and found that the distance correlation strategy indeed performs well. Furthermore, most previous studies did not allow for natural churn in the portfolios, the presence of which we believe makes distance correlation an even more attractive option. Thirdly, as mid-cap companies are more illiquid compared to large-caps, their available trading history may be shorter, less uniform, or missing for some period of time. The distance correlation measure can handle such time series features and still produce a quantitative measure of the relationship between firms. This allows us not to drop any firms, thus avoiding selection bias. Hence, we use the covariance matrix generated by distance correlations to produce a fully connected initial network of firms.
Subsequently, we apply the Triangulated Maximally Filtered Graph (TMFG) method introduced by \citet{Massara2017} to filter this dense matrix. This process results in a network with fewer edges, representing the strength of the relationships between firms with minimal loss of information. The added sparsity yields more meaningful initial relationships that subsequent models can build upon. 

Once networks are formed and stock price data are included, traditional methods fall short on handling the complex resulting data structure. Deep learning techniques, however, with their ability to create higher-order representations of any available data, commonly excel at this. As they do not impose restrictions on the data distribution and can handle non-standard data types by design, they are applicable in a wide variety of settings. For example, they have produced state-of-the-art results in several domains, such as speech recognition, natural language processing, object detection, drug discovery, and genomics \citep{LeCun2015}. Furthermore, deep learning models tend to scale well to high-dimensional datasets, \citet{avramov2023} applied deep learning techniques such as feed-forward neural networks and conditional autoencoders to identify mispriced stocks consistent with most anomaly-based trading strategies. A recent study on midcap default prediction has also shown \citep{Korangi2022},  they can be designed to find optimal solutions for various problem types with different objectives or constraints. This suggests that deep learning methods may be well-suited to mid-cap portfolio optimisation.

In this paper, we propose employing a class of deep learning methods, Graph Neural Networks (GNN), which can learn non-linear, complex representations of the firms. More specifically, we use Graph Attention networks (GAT), a variant of GNNs that employs attention mechanisms to weigh the importance of a firm's neighbouring nodes \citep{Velickovic2018}. Unlike previous topological studies, which have shown that network structure can play a vital role in portfolio optimisation, GNNs distil information from the relationships to produce the portfolio weights without relying on a few static measures. GATs, in particular, work on specific sub-structures of graph data, while the graph data can be dynamic. This makes them an attractive option for portfolio optimisation, as the relationships between firms can change over time and may behave differently under different macroeconomic environments. The optimisation procedures of deep learning architectures also work well in higher-dimensional space, such as the historical returns of a large number of firms. Like other deep learning architectures, GATs can also be trained to optimise any chosen objective function and are flexible in terms of the types of output that can be produced.

To measure portfolio performance, we use the widely applied Sharpe ratio, i.e.\@ the ratio of returns over volatility \citep{Sharpe1966}. We employ a similar criterion for model training, by using a custom loss function derived from the Sharpe ratio. This allows us to target generating optimal portfolio weights, without having to predict individual returns in the same way that, for example, \citet{Zhang2020} did. To cope with the dynamic nature of the problem and allow the set of active mid-cap firms to vary over time, we use a rolling window approach which allows graph inputs and past returns data to vary as we move between forecast periods.

Therefore, the three key research questions addressed in the paper are the following: 
\begin{enumerate}
\item Can an effective network topology be constructed from sparse historical data on a large collection of firms?
\item Are graph attention networks able to generate higher-order representations of this network that enable constructing an optimal portfolio for mid-caps?
\item How does the model perform under different market conditions, and can we infer useful strategies from the model results?
\end{enumerate}

In so doing, the paper makes three main contributions. First, we develop topological portfolio optimisation models, extending the literature that hitherto focused chiefly on stochastic models, and applying the resulting approach to the challenging set of mid-cap firms. Second, by using GPUs and parallel computing, we are the first to be able to apply the distance correlation measure and subsequent TMFG filtering at this scale. Finally, we use the resulting networks as inputs to graph-based deep learning models and show how these are capable of producing portfolio weights for the large number of firms we are faced with.

The remainder of the paper is organised as follows. Section~\ref{sec:Literature Review} reviews the relevant literature, focusing on large-scale portfolio optimisation, graphs and graph-based deep learning models. Section~\ref{sec:Methods} describes the data and the process by which this data is converted into graphs, as well as defining several measures used in our work. The proposed models, and the baseline models against which they are compared, are further described in Section~\ref{sec:Models2}.  
Section~\ref{sec:ExperimentalSettings2} elaborates on how we set up the empirical analysis, summarising the different steps and types of model comparisons made. Section~\ref{sec:Results2} then presents and discusses the results. Finally, Section~\ref{sec:Conclusion2} summarises the main insights gained from our study and suggests some future research.

\section {Related literature} 
\label{sec:Literature Review}
In this section, we discuss relevant literature for the study, focusing on large-scale portfolio optimisation studies and GNNs.

\subsection{Correlation networks and portfolio models}
\label{subsec:Lit_review_part_one}
Portfolio optimisation has been the subject of a large body of research, and novel methods, with various constraints and objectives, continue to be developed \citep{DeMiguel2009, Branch2019}. Of particular interest to our work are studies that have focused on optimising large-scale portfolios effectively. \citet{Perold1984} was the first to do so, by considering the specific nature of dense covariance matrices, and recommending strategies to make them sparse so that their analysis becomes computationally feasible. More broadly, the first branch of large-scale portfolio optimisation studies followed a similar approach by devising algorithms to reduce computational time and memory space requirements for the classical mean-variance approach.

Later studies on large-scale portfolio optimisation have focused on proposing model extensions and computational methods to solve them, measuring the performance of the resulting allocation weights using performance metrics such as the Sharpe ratio. Extending the mean-variance framework by adding a probabilistic constraint requiring that the expected returns exceed a chosen threshold with a high confidence level, and introducing additional trading constraints, \citet{Bonami2009} proposed a novel exact solution for their resulting model, which they tested on a portfolio of up to 200 firms. Demonstrating their approach for pools of stocks of up to 100 S\&P 500 firms, \citet{Ao2019} proposed an unconstrained regression representation of the mean-variance portfolio problem, which they estimated using sparse regression techniques. \citet{Bian2020} and \citet{Dong2020} used regularisation methods for portfolio optimisation as, without such methods, a large universe of stocks would lead to overly small or unstable allocations (and, hence, high transaction costs). They found that these techniques improved portfolio performance (in terms of the Sharpe ratio) compared to the standard model. Performance-based regularisation, whereby the sample variance of the estimated portfolio risk and return is restricted, also performed better on several Fama-French data sets \citep{ban2018}.  Recently, \citet{Bertsimas2022} reviewed the size of portfolios that previous large-scale portfolio optimisation studies were able to handle and proposed a ridge regression-based regularisation algorithm that speeds up the convergence of sparse portfolio selection. They showed that their method can select up to 1000 stocks from the Wilshire 5000 equity index. However, they did so without reporting the Sharpe ratio of this selection.

Despite these computational advances, we argue that, in all of the aforementioned work, either there was further scope for increasing the portfolio size, or the performance analyses to measure the efficacy of the proposed algorithms were restricted to using simulated data. Additionally, in all of these studies, the universe of firms or assets from which to select was always kept constant, and firms that default or those that are acquired or liquidated were thus omitted, eliminating an important driver of idiosyncratic portfolio risk. Instead, we take a dynamic approach and allow both the universe of selectable stocks and the chosen portfolio to change over time. In so doing, we avoid selection bias and ensure that our approach more closely mirrors a real-world scenario wherein investors wish to invest in a certain market sector or asset class. We seek to make an optimal decision considering all of the firms available within that asset class at any given time. 

In order to create a sparse covariance matrix, another alternative explored in the literature is information filtering using graphs or network data. How to build sparse networks that represent information contained in large data sets is an active area of study in various domains such as internet search \citep{xie2018}, social networks \citep{Berkhout2019} and, similarly to our paper, finance \citep{fan2013}. New applications also continue to emerge, such as social network analysis for link prediction \citep{Zhang2018a}, recommender systems \citep{Fan2019}, or the study of object interactions in complex systems \citep{Battaglia2016}. In previous work related to ours, starting from a graph representing correlated assets, \citet{Onnela2003} and \citet{cho2023} showed that investing in the peripheries identified by a filtered subgraph provided benefits for portfolio diversification. They used a static slice of S\&P 500 companies, starting at a larger base than previous studies, but the firms again remained unchanged over the long time frame of 20 years that they studied. 

In any of these application settings, graph filters aim to maintain the most relevant information by constraining the topology of the graphs. For example, identifying the Minimum-Spanning Tree (MST) is a filtering mechanism for dense graphs that keeps the edges with the highest weights and allows no cycles or loops in the graph \citep{Mantegna1999}. The Planar Maximally Filtered Graph (PMFG) imposes a different constraint on the graph's topology, requiring it to be planar; i.e., there should be no edge crossing on a plane \citep{Tumminello2005}. Compared to MSTs, PMFGs were found to be more robust for financial market networks as market conditions change, without losing much information content \citep{Yan2015}. Alternatively, Triangulated Maximally Filtered Graph (TMFG) is a more computationally efficient algorithm since, unlike PMFG, it can be parallelised \citep{Massara2017}. In this work, we adopt the latter method for correlation networks of stocks, making them suitable for the large datasets we work with.

As we mentioned earlier, we use the distance correlation measure to account for the strength of the relationship between firms. Alternatively, \citet{Diebold2012,Diebold2014} developed the connectedness metric, using VAR (Vector Auto Regression) decomposition methods, to measure the pairwise relationship between firms. Their approach can identify individuals or clusters of firms crucial to networks and quantify the direction of risk spillovers. However, these methods have limitations for large networks, such as those for mid-cap firms, due to the amount of historical data they require. Given these limitations, we instead used the distance correlation measure for pairs of firms, and TMFG for filtering, to provide the sparse network which serves as the input to our deep learning-based models. For a more extensive survey encompassing time-series correlations and network filtering in financial markets, we refer the reader to \citet{marti2021}.

\subsection{Graph Neural Networks}
\label{subsec:Lit_review_part_two}
GNNs were first proposed by \citet{Scarselli2009} for node classification tasks. Similarly to recurrent neural networks (RNNs), the first generation of GNNs employed recursion, which they used to learn higher-order representations for a node from its neighbours. As the deep learning field evolved with the emergence of RNNs for sequential data, Convolutional Neural Networks (CNNs) for primarily image processing, and attention-based models \citep{Vaswani2017} for spatial analysis of unstructured data, non-euclidean data models based on GNNs also developed in parallel. Graph Convolutional Networks (GCNs) borrowed concepts from CNNs, such as kernel filter size and stride, to generate representations for graphs \citep{Kipf2017}. They produced state-of-the-art results on popular graph datasets such as citation networks and knowledge graphs, outperforming semi-supervised or skip-gram-based graph embeddings, label propagation and regularisation approaches. Graph Attention Networks (GATs) further improved on these results by introducing variable aggregation of neighbours, and they also proved successful in transductive learning tasks where the data is not fully labelled \citep{Velickovic2018}. We refer the reader to a comprehensive survey by \citet{Wu2021} for a general introduction to GNNs and their various flavours. 

Thus far, GNNs have been applied in some application areas related to finance or financial markets. For example, they were used in consumer finance for fraud detection \citep{Xu2021} and credit risk prediction \citep{oskarsdottir2021}. \citet{feng2022} used a combination of a GCN model and self-attention (unlike in a GAT, which has self-attention as part of its own architecture) for making stock recommendations, i.e.\@ to predict the top 3 stocks (out of 738 stocks) for the next period. 

Overall, we argue that deep learning methods, such as GATs, are better deployed at scale, on a large set of illiquid, risky firms such as mid-caps, than on a small set of established firms, as they are well equipped to extract complex relationships but need a large enough amount of data to produce stable solutions.

\section {Methods} \label{sec:Methods}
\vspace{+1pt}
In this section, we describe the data used for this study, the distance correlation measure and the graph filtering algorithm, which provide the inputs for the GNN and other benchmark models. 

\subsection{Data}
\begin{figure}[h]
\includegraphics[width=\linewidth]{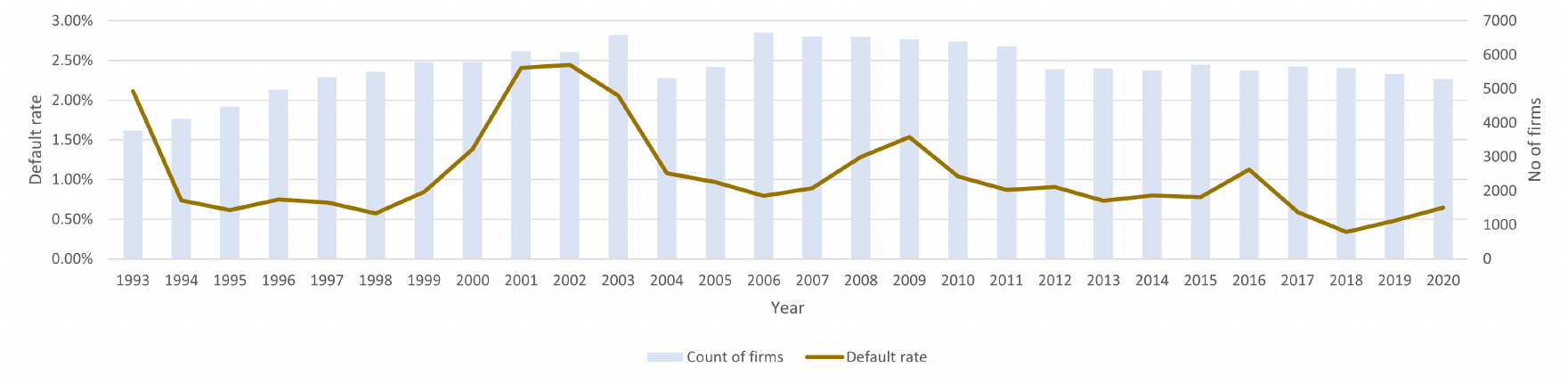}
\centering
  \caption{Number of firms and year-on-year default rate in the sample}
   \label{fig: Default evolution}
\end{figure}

We collected the daily closing prices of all mid-cap companies listed in the US over 30 years, from 1990 to 2021. A total of 16,793 firms were active for at least part of this period. For portfolio selection, we use a three-year rolling window, which limits the allocation problem to around 5,000 firms at any given time point. This number changes substantially over time due to defaults and firms entering or leaving the mid-cap universe, as can be seen on a year-by-year basis from Figure~\ref{fig: Default evolution}. Here, default is when a firm has entered liquidation/bankruptcy or experienced another credit event that adversely affected the firm's equity. The default rate varied around 1.5\% in any given year; over the full period, 8.5\% of total firms faced some form of default.

We convert the prices to daily returns for each firm, denoted by $r_{ut}$ for a firm $u$ at time $t$. These return series are divided over time into a training (50\%), validation (25\%), and test set (25\%), with the most recent data serving as the test set. We have more details of this split in Section~\ref{subsec:overallarchitecture} 
To enable capturing relationships between firms, we calculate the return-volatility series using the standard deviation of these returns and a 30-day lookback period. Using return-volatility series is in line with previous work on connectedness by \citet{Diebold2012, Diebold2014}, who, like us, employed return volatilities instead of returns to look for such correlations. In financial market settings, the return-volatility series has some interesting properties compared to the return series, such as exhibiting pronounced co-movements during short risk-off periods but showing weaker relationships under more benign market conditions. This series also tends to have strong serial correlation and closely mirrors market investor sentiment (e.g.\@ being positively correlated with the volatility index, VIX), which are essential for identifying crises \citep{huang2019}. Those are precisely the circumstances in which the performance of different portfolio allocation strategies can diverge substantially.

More formally, for a firm $u$ at time $t$, return volatility is defined as $c_{ut} = \sigma(r_{ut},r_{ut-1},\ldots,r_{ut-29})$. 
We let $\mathcal{V}$ represent the total universe of firms, which we observe over $N$ periods, with $ \mathcal{V}_{t} \subseteq \mathcal{V} $ denoting the set of firms active at time $t$ ($t = 1,2,\ldots,N$). Using a lookback period of $T$ (set to 3 years)
, each firm $ u \in \mathcal{V}_{t} $ has a daily return series, 

    \begin{equation} \label{eq:returns}
  x_{ut} =(r_{ut-1},r_{ut-2},\ldots,r_{ut-T}) \in \mathbb{R}^{T},
\end{equation} and a daily return-volatility series,  
\begin{equation} \label{eq:vol}
 l_{ut} =(c_{ut-1},c_{ut-2},\ldots,c_{ut-T}) \in \mathbb{R}^{T} 
.\end{equation}

To prepare the input to the models, we stack the individual return series from \eqref{eq:returns} for all firms in $\mathcal{V}_{t}$, to obtain the feature matrix $X_{t} \in \mathbb{R}^{|\mathcal{V}_{t}| \times T}$, i.e. 

\begin{equation} \label{eq:featureX}
     X_{t} =[x_{1t} \; x_{2t} \, \ldots \, x_{ut} \, \ldots \, x_{|\mathcal{V}_{t}|t}]^{\top} 
.\end{equation}

\subsection{Distance correlation}\label{subsec:DistanceCorrelation}
We quantify the strength of relationship between two firms by evaluating their return volatilities and applying the distance correlation measure. 
Distance correlation is a generalised measure of dependence, which is capable of capturing non-linear dependencies and is known to perform well in domains such as signal processing \citep{Brankovic2019} and computational biology \citep{Mendes2018}. 
Starting from the volatility series $l_{u}$, $l_{v}$ for two firms $u$, $v$ (see equation \eqref{eq:vol} but omitting $t$ for brevity)
, our distance correlation measure, $dcor(u,v)$, is derived as follows. 

For each firm, we consider the absolute change in volatility between any times $i$ and $j$ over some (lookback) period of length $T$, and then double-centre the resulting $T \times T$ matrix. Each such rescaled firm-level change matrix can now be compared against the matrices of the other firms, to derive the distance correlation between each pair of firms. 
More formally, we first define two matrices $A=(a_{i,j})$ and $B=(b_{i,j})$, for a pair of firms $u$ and $v$, respectively, as  

$$ a_{i,j} = ||l_{ui}-l_{uj}|| $$  
$$ b_{i,j} = ||l_{vi}-l_{vj}|| $$ where 
$i,j \in \{{t-1,t-2,...,t-T}\}$, 
and $||\cdot||$ is the Euclidean distance.  Each such matrix captures times when the corresponding firm has higher or lower volatility compared to other time periods. By doing so across all times, we obtain quantified values of the firm's volatility changes over the observed time period. This sets up a comparison with another firm over the same period. To make the values comparable between two firms with different risk characteristics, we define two further matrices $A^{'}=(a^{'}_{i,j})$, $B^{'}=(b^{'}_{i,j})$, $i,j=1,2,...,T$, that normalise the matrices $A$, $B$, 
  
$$ a^{'}_{i,j} = a_{i,j}-a_{.j}-a_{i.}+a{..} $$ 
$$ b^{'}_{i,j} = b_{i,j}-b_{.j}-b_{i.}+b{..} $$ where $a_{.j}$ is the mean across rows, $a_{i.}$ is the mean across columns and $a_{..}$ is the mean across all values in matrix $A$ (and similarly for $B$). The distance covariance, $dcov(u,v)$,
now is the average over all entries of the element-wise multiplication of $A^{'}$ and $B^{'}$, from which the distance correlation can then be obtained, as follows: 
  $$dcov(u,v) = 1/T^{2}\sum_{i=t-1}^{t-T}\sum_{j=t-1}^{t-T}
  a^{'}_{i,j} \,
  b^{'}_{i,j}
  $$
  \begin{equation} \label{eq:dcor}
        dcor(u,v) = dcov(u,v)/\sqrt{dcov(u,u) \, dcov(v,v)}.
  \end{equation}
This measure has some valuable properties that are relevant to our problem \citep{Szekely2007}: 

\begin{enumerate}
  \item $dcor(u,v)=0$, if and only if $l_{u}$ and $l_{v}$ are independent.
  
  \item $0 \leq dcor(u,v) \leq 1$, unlike Pearson correlation which instead captures the linear dependence as a number between -1 and 1. This is useful because we are interested in the strength of the dependence rather than the direction of the dependence.
  
  \item The measure can produce a value for two series of unequal length. Given that firms can drop in and out of the universe, resulting in different histories, this feature allows one to nonetheless consider the relationship between them.
\end{enumerate}

A naive implementation of distance correlation calculation has $O(T^{2})$ time complexity for a pair of firms. Optimisation methods exist to implement this in $O(T \log T)$ when faced with two series. Still, the main problem is to calculate correlations between many firms and over multiple periods \citep{huo2016}. In our experiments, we chose to implement the required computations in a distributed architecture, using a mix of GPUs and CPUs. Parallelising the pairwise comparisons between firms (using workload manager SLURM and High Performance Computing) facilitates the calculation of large correlation matrices. The resulting code can handle the problem size in a reasonable time. 

Writing $d_{i,j}$ as a shorthand for the distance correlation, $dcor(v_i,v_j)$, computed between any two nodes $v_i, v_j \in \mathcal{V}_{t}$, the resulting dependency matrix, $D_{t} = (d_{i,j})$, generally leads to a complete graph in which each firm is connected to all other firms. 

Some studies have used threshold conditions to remove weaker edges (those associated with small correlation values) from this graph, but the choice of threshold value is arbitrary, and the composition of our portfolio changes over time, further complicating the use of a global threshold mechanism. Instead, we use a more advanced technique from graph theory to filter $D_{t}$ and remove weaker connections from it. 

\subsection{Graph filtering}\label{subsec:GraphFiltering} 

The filtering technique we chose is the Triangulated Maximum Filtered Graph (TMFG) method proposed by \citet{Massara2017}, which, like PMFG, imposes a planarity constraint on the graph but is more scalable to larger datasets such as ours. Using the topological features of the graph as a constraint, a planar graph retains most of the information with fewer edges. A planar graph can be drawn on a plane (or a sphere) without any two edges crossing. Such graphs have attractive features making them tractable for analysis, by, for example, simplifying cluster or community detection.

We denote by $\mathcal{K} = (\mathcal{V}_{t},\mathcal{F}_{t})$ the dense graph before filtering, wherein, for any two nodes $v_i, v_j \in \mathcal{V}_t$ ($i \neq j$), we let $(v_i, v_j) \in \mathcal{F}_{t}$ if and only if $d_{i,j} > 0$. 
Using the distance correlation values from Equation \eqref{eq:dcor} as edge weights, TMFG 
filters $\mathcal{K}$ by searching for a (near-)maximal planar subgraph $\mathcal{G}
$, i.e.\@, one with the highest possible sum of retained edge weights. Planarity constraints reduce the edges from $|\mathcal{V}_{t}|(|\mathcal{V}_{t}|-1)/2$ in $\mathcal{K}$ to at most $3(|\mathcal{V}_{t}|-2)$ in graph $\mathcal{G}$. 

The TMFG algorithm grows this planar graph by optimising a chosen score function at each iteration step. In this paper, we select as score function the sum of edge weights between pairs of firms, as given by our distance correlation measure.  The procedure starts by identifying a clique of four firms that have the largest such sum of correlations among all firms. A clique is when all the distinct vertices in the sub-graph have an edge between them, i.e, all clique members are connected. This happens in most cases in our volatility networks, as all firms are correlated and unlikely to have a zero value. 
Next, out of all remaining firms, the algorithm looks for the node (firm) that has the largest sum of correlations (here: distance correlations) with any connected subset of three nodes that are already in the network (so, initially, three within that clique of four) and extends the network with this node and the extra three connections.  It keeps repeating this until all firms are added again to the network. Doing so guarantees that the new graph is planar and sparse.
To improve computational efficiency, the algorithm maintains and updates incrementally a cache of possible combinations of these sub-graphs. At the end of this process, a sub-graph $\mathcal{G} = (\mathcal{V}_t, \mathcal{E}_t)$, in which $\mathcal{E}_t \subseteq \mathcal{F}_t,$ is created. 
 
\begin{figure}[h]
\includegraphics[width=\linewidth]{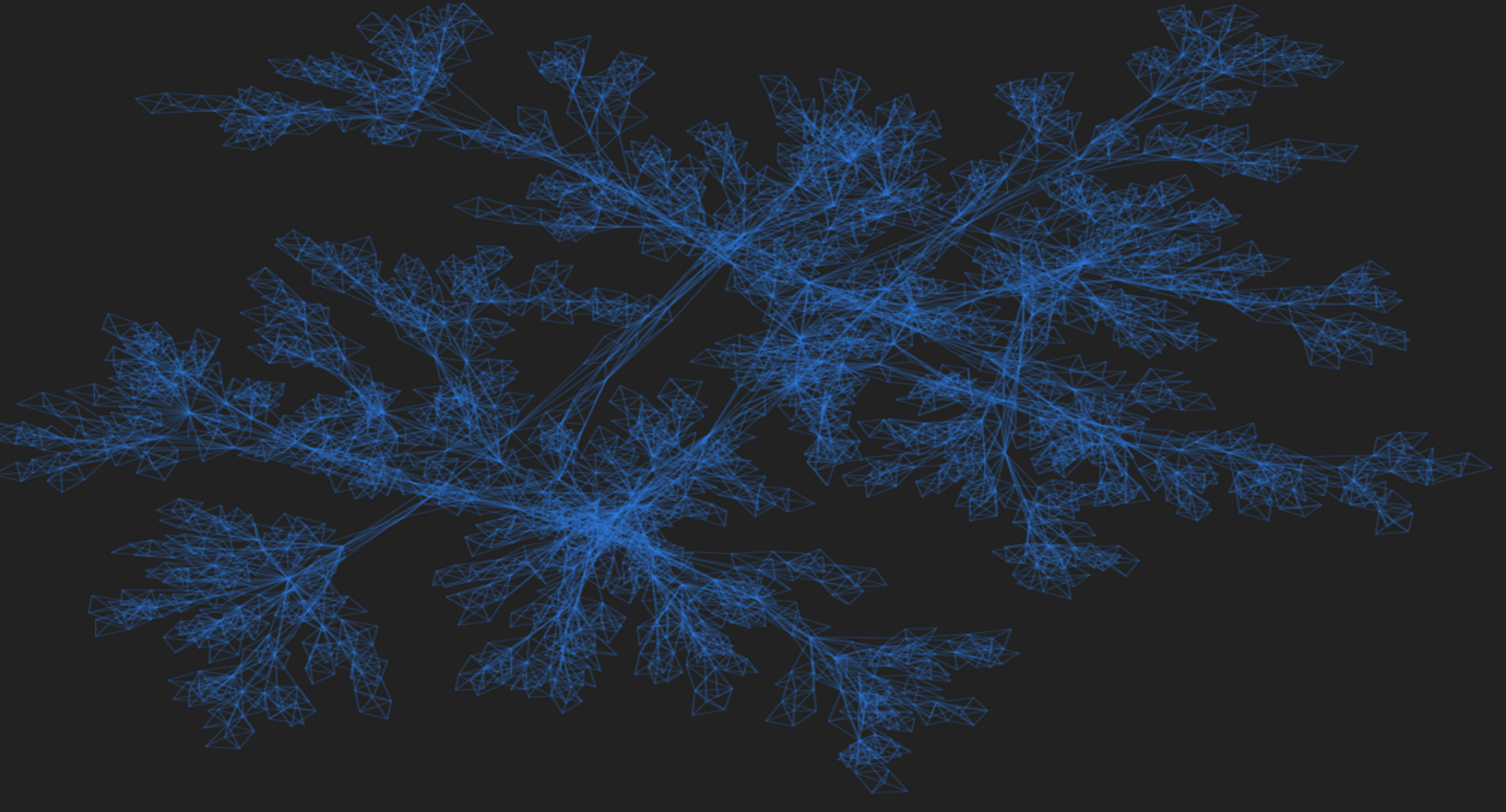}
\centering
  \caption{Snapshot of network of mid-cap firms showing the remaining node connections after TMFG filtering.}
   \label{fig: Full Graph}
\end{figure}

The resulting graph will serve as input to the deep learning models and graph-based portfolio models. Its planarity property also aids visual representation. Figure~\ref{fig: Full Graph} depicts a sample graph extracted from our data. This shows, for an example time period, the topology of the generated network as being characterised by a fairly small number of central nodes and a large periphery. As we roll over the data window, we construct such a graph for each subsequent time period.

\subsection{Network measures}\label{subsec:Network Measures}
The graphs created after TMFG filtering could be used directly for portfolio optimisation, as some studies \citep{Li2019,pozzi2013a} have done. Rather than employing deep learning in a subsequent step, these prior approaches basically involve investing in the graph's most `peripheral' assets. As we believe it is helpful to compare our new GAT-based approach against such a simpler network-index based approach, we next outline an inverse peripherality score, $p_{i}$, for each node (firm) $i \in \mathcal{V}_{t}$ in graph $\mathcal{G}$, which will later be used in section~\ref{subsec:Network index model} to determine the allocation weights for this benchmark model. 

Specifically, we propose a composite score incorporating three common centrality measures: a node's degree, betweenness and closeness. These metrics have been studied since the 1970s \citep{Freeman1977} and remain popular in the network analysis literature \citep{oskarsdottir2021}. For the first of these, i.e.\@ the degree centrality of node $i$, we take the number of nodes to which $i$ is directly connected (also referred to as the node's neighbours) relative to the total number of nodes (other than $i$ itself). More formally, having defined the set of neighbours of $i$ as  

   \begin{equation} \label{eq:neighbours}
       \mathcal{N}_{i} = \{{v | (i,v) \in  \mathcal{E}_{t}}\}
   \end{equation} 
the degree centrality, $dc_{i}$, for firm $i$, is defined as 

\begin{equation}
  dc_{i} = | \mathcal{N}_{i}|/(|\mathcal{V}_{t}|-1)
\end{equation}

The second measure, betweenness centrality, can be thought of as an indicator of the amount of activity that passes through a graph node when any changes occur in the network. To measure this activity level, it considers the shortest paths between all pairs of nodes. The betweenness centrality of a node is the fraction of these shortest paths (other than those starting or ending in the node) that pass through that node.

For a firm $i$, this is defined as
\begin{equation} \label{eq:bc}
    bc_{i} = \sum_{u,v \in \mathcal{V}_{t}}\frac{s(u,v|i)}{s(u,v)}
\end{equation}

\noindent where $s(u,v)$ is the number of shortest paths between $(u,v)$, and $s(u,v|i)$ is the number of these shortest paths that pass through $i$, with $s(u,v|i) = 0$ if $i = u$ or $i= v$, and $s(u,v) = 1 $ if $u=v$.

Our third centrality measure is closeness centrality, $cc_i$, which, for a given node $i$, considers the reciprocal of the average length of the shortest paths to all other nodes that are reachable from that node. 
Hence, for a firm $i$,
\begin{equation}
    cc_{i} = \frac{(|\mathcal{B}_{it}|)}{(|\mathcal{V}_{t}|-1)} \frac{(|\mathcal{B}_{it}|)}{\sum_{u \in \mathcal{B}_{it}}d(u,i)}
\end{equation}

\noindent where $\mathcal{B}_{it}$ contains the set of nodes that one can reach from $i$, not including $i$ itself, and $d(u,i)$ is the shortest distance (in terms of edge count) between $(u,i)$; for example, 
$d(u,i)=1$ if there is an edge that directly connects both nodes. In our setting, the TMFG graph ($\mathcal{G}$) still connects all the nodes, so all nodes are reachable from one another, but, unlike in $\mathcal{K}$, the distance now varies. As with the other two measures, higher scores for $cc_i$ imply higher centrality, with values ranging between zero and one.

We use the networkx library implemented by \citet{hagberg2008a} to calculate these three centrality measures. Averaging them produces a simple inverse peripherality (centrality) score, $p_i$, for each firm $i$,
\begin{equation}
    p_{i} = (dc_{i}+bc_{i}+cc_{i})/3.
\end{equation}    

\subsection{Model performance and loss metric}\label{subsec:Sharpe Ratio }

We adopt the Sharpe ratio as the final performance measure for the models \citep{Sharpe1966}. This is a well-studied metric to measure portfolio performance. Whilst there have been studies proposing further refinements to deal with some of its limitations 
\citep{Lo2002,Farinelli2008}, we use the widely accepted form of the metric. 
For an individual firm $u$, we estimate the Sharpe ratio from the sample mean and variance of the returns (defined in \eqref{eq:returns}). The same method also applies to portfolio returns. To produce the latter, we take a weighted sum of the individual returns using the corresponding allocation weights in the portfolio, giving a return series that has a similar format to an individual firm's return series. 

Thus, for a series with $T$ daily returns, a firm's mean return and return variance are given by: 
 
$$ {\mu}_{u} = \frac{1}{T} \sum_{t=1}^{T} r_{ut} $$ 

$$ {\sigma_{u}^{2}} = \frac{1}{T} \sum_{t=1}^{T} (r_{ut}- {\mu}_{u})^{2}. $$ 

From these quantities, the Sharpe ratio, ${SR}_{u}$,
can be easily estimated as 

$$ {SR}_{u} = \frac{{\mu}_{u} - r_{f}
}{{\sigma_{u}}} $$

\noindent where $R_{f}$ is the risk-free interest rate. For ease of calculation, we set this rate to zero  as a constant baseline for all. Therefore, the Sharpe ratios produced here cannot be compared to those reported by other studies, but they do allow for a direct comparison between the models in this study. After training, we use them to test model performance and report how they evolve. For the model training itself, we take a different approach outlined below.
    
Most supervised deep learning models have a prediction target, but in our problem, we do not aim to predict but provide a score for each firm, i.e.\@ the weight to be assigned to that firm in the portfolio. To enable this, we can convert the Sharpe ratio to a suitable loss metric, which the deep learning models will seek to minimise during training. Maximising the Sharpe ratio is equivalent to minimising the negative logarithm of that ratio, which gives us a more convenient loss function. A similar approach was used by \citet{Zhang2020}, using gradient ascent and a differentiable Sharpe ratio. The loss function is thus expressed for a portfolio $p$ with a daily returns $r_{pt}$ at time $t$ as 

\begin{equation}
     LF = -\ln({\mu_{p}}) + \ln({\sigma_{p}}).
\end{equation}

where $r_{pt}$ is calculated from the weights vector $W_{t} \in \mathbb{R}^{|\mathcal{V}_{t}|}$,  
\begin{equation}
    r_{pt} = \sum_{i=1}^{|\mathcal{V}_{t}|} w_{i}r_{it}
\end{equation}

\section{Models}\label{sec:Models2}
\vspace{+5pt}

\subsection{Graph Attention Networks} \label{subsec:Graph Attention Networks}

Different types of GNNs have been developed to learn from graph data. In this paper, we opted for GATs rather than the earlier class of GCNs. As explained in section \ref{subsec:Lit_review_part_two}, the latter were introduced along with convolutional neural networks, which were designed for image processing. Unlike image data, however, graph data are more complex in that they may have a variety of features and node connections that vary in importance. GCNs generate a higher-order representation of input features and neighbours, by weighting the features of each neighbour based on its respective degree centrality. This choice of a single measure rules out more complex weighting mechanisms.  
GATs solve this problem by using the self-attention mechanism. The latter introduces learnable parameters to generate the weights for neighbours, making GATs more flexible in how they learn from the neighbours' features \citep{Velickovic2018}. This is particularly attractive in many real-world settings where, like in ours, the graphs are dynamic and evolve as time passes, or in many financial settings, where market conditions also tend to vary over time. In those settings, fixed weighting of neighbours might not perform well. The mechanism also allows for different features to be learnt through multiple heads, such as short-term moves in one head and longer-term relationships in another. Furthermore, the attention operations are more efficient than alternative approaches, since they are parallelisable across node neighbour pairs. 

Here, we formally define the GAT specific to our problem. Given a graph $\mathcal{G}= (\mathcal{V}_{t},\mathcal{E}_{t})$ as defined in section \ref{subsec:GraphFiltering}, and the input features defined in \eqref{eq:featureX}, GAT transforms the input features $X_{t}$ into a higher-order representation $H_{t}$ given by

 \begin{equation} \label{eq:H}
     H_{t} = [h_{1} \; h_{2} \, \ldots \, h_{u} \, \ldots \, h_{|\mathcal{V}_{t}|}]^{\top} 
 \end{equation}
\noindent where $h_{u} \in \mathbb{R}^{T'}$

Note the dimensionality change from $T$ to $T'$ with this transformation. Specifically, for a given firm $u$ (and, from here on, omitting the current time $t$ for brevity), the transformation function from $x_{u}$ to $h_{u}$, for a GAT with $K$ heads, is defined as 
 \begin{equation} \label{eq:gath}
     h_{u} = \parallel_{k=1}^{K} F(x_{u},\Sigma_{v \in {N_{u}}}a_{k}(u,v)(W_{k}x_{v})) 
 \end{equation} 
in which $N_{u}$, as before, are the neighbours of firm $u$,. For each head $k$,  $W \in \mathbb{R}^{T' \! \times T}$ is a weight parameter matrix of the model that is learnt during training, $a(u,v)$ is the weighted importance score of adjacent firm $v$, $F(\cdot)$ applies an activation function (ReLU), and $||$ is the concatenation operator applied to the outputs of all $K$ heads.

The function $a(u,v)$ is where each type of GNN differs; whereas it was a convolutional function for GCNs, it is attention for GATs,
 
  \begin{equation}
      a(u,v) = softmax\left( \sigma( a 
      \left[Wx_{u} || Wx_{v}\right])\right)
  \end{equation} 
where $\sigma$ is a non-linear function (specifically, Leaky Regularized Linear Unit, in short LeakyReLU), $a \in \mathbb{R}^{2T'}  $ is a weight vector, and $||$ is again the concatenation operator. In our experiments, we set $T'=24$ and $K=8$.

Above, $H_{t}$ is a higher-order representation of the input features $X_{t}$. To convert this representation to a one-dimensional portfolio weights vector, we next reduce the dimensionality by adding a series of learnable layers. 
We use two blocks of feed-forward networks, which first apply batch normalisation. This normalises the feature inputs in the standard way, using mean and standard deviation, which enables faster convergence and optimisation. We pass this through the first feed-forward network and then apply dropout. The dropout process helps to improve the stability of the training, reducing overfitting. The second feed-forward network uses L1, or LASSO, regularisation to further shrink and eliminate unnecessary weights. We denote the output of each layer $i$ by $s_{iu}$.

$$s_{iu} = \sigma^{'}(W_{iu}*h_{u} + b_{i}) $$ where  $i=1,2$ are the two feed forward networks, $W_{i}$ and $b_{i}$ are their trainable weights and bias terms, respectively, and $\sigma^{'}$ is a nonlinear function (for which we used ReLU). The $s_{1}$ passes through a normalization layer and dropout layer before it reaches the next feed-forward block. The final output $s_{2} \in \mathbb{R}$ provides the (as yet unscaled) scores for the starting universe of firms.

\begin{figure}[htbp]
\includegraphics[width=\textwidth]{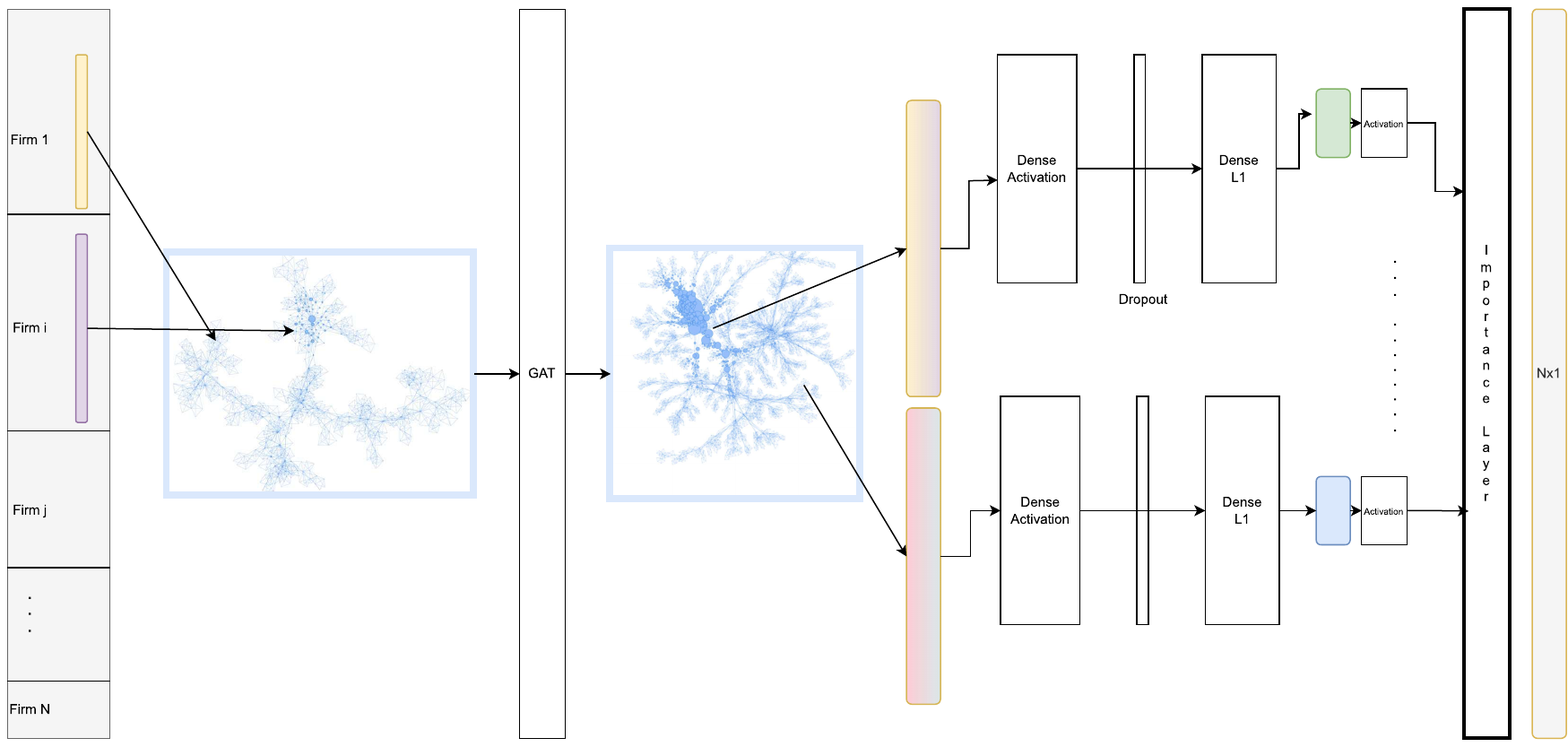}
\centering
  \caption{GAT-based model: First embeddings are obtained from GAT and for each node new representations are created before they are combined to form portfolio weights}
   \label{fig:Gat model}
\end{figure}

We introduce a final allocation layer in the model to rescale these scores to generate the final weights and meet the weight constraints, i.e.\@, making sure that the values range between $[0,1]$ and add up to $1$. This layer is labelled `Importance Layer' in Figure~\ref{fig:Gat model}. A softmax output function would have been the standard solution to generate these weights, as used by most deep learning papers if they need the outputs to add up to one. However, given the size of our investment universe, softmax would lead to many tiny holdings of 
firms, which is impractical and can bring high transaction costs. Instead, we prefer to concentrate the portfolio in fewer positions, which reduces the cost of managing the portfolio. To this end, we considered two alternatives. One is to use sparsemax, which generates sparse outputs that would be more suitable to our application scope \citep{Martins2016}. Another is to introduce a weight reduction mechanism in the final layer of the model, that generates the final weights and uses the output of the feed-forward network $s_{2}$ to bring the sparsity with regularisation, thus bringing the weights to what we need and with allocations in fewer firms,

  $$ w_{u} =  \frac{s_{2u}}{\sum_{v=1}^{|\mathcal{V}_{t}|} s_{2v}}.$$
  
We found our mechanism more stable and straightforward to implement than softmax and sparsemax. The training loss function reduction tended to be smoother and more consistent, with the same input producing similar weights over different runs. In contrast, the convergence path with the other training mechanisms was noisier, 
i.e.\@ the loss over the epochs was more volatile. Note that this allocation layer could be easily extended to meet other portfolio constraints (e.g.\@ choosing the top $K$ firms).

We use standard GATs with their original parameters where possible and implemented them using the Spektral package developed by \citet{Grattarola2020}. The full model with the added layers is summarised in Figure \ref{fig:Gat model}. The data embeddings are displayed in colour and the deep learning layers are shown as unfilled boxes in the figure. The GAT layer takes a graph as input. Each node in the graph also has the corresponding firm's return series as node features. For each of these nodes, the GAT model generates an embedding which is further processed by dense layers with non-linear (ReLU) activation, dropout and L1 regularisation, as previously discussed. These scores are then converted to portfolio weights in the importance layer, which collects the scores from earlier steps and, using the reduce-weight mechanism previously described, allocates weights. From this, we can obtain a series of Sharpe ratios for each graph, which we then average to measure the model's performance over the full time period. 

\subsection{Benchmark portfolios} \label{subsec:Benchmarkportfolios}

\subsubsection{Mean-variance model} \label{subsec:Markowitz Portfolio} \hfill\\
\indent As discussed earlier, \citet{Markowitz1952}'s mean-variance model is widely recognised as a cornerstone of modern portfolio theory. Therefore, we include the model as one of the benchmark models. For $|\mathcal{V}_{t}|$ firms active at time $t$, the model assists in determining the optimal weights $w_i$ for each asset $i$ in the portfolio, thereby requiring that $\sum_{i=1}^{|\mathcal{V}_{t}|} w_i = 1$. In so doing, one looks for an optimal trade-off between the expected (here, quarterly) returns and the portfolio's volatility. 

In its classical implementation, the expected return and the variance of the portfolio, $E(R_p)$ and $V(R_p)$, are estimated from the sample using the sample mean and covariance,  
\begin{equation}
    E(R_p) = \sum_{i=1}^{|\mathcal{V}_{t}|} w_{i}\mu_{i}
\end{equation}
\begin{equation}
    V(R_p) = \sum_{i}\sum_{j}w_{i}w_{j}cov(i,j) 
\end{equation}
where $\mu_{i}$ is the mean return of $i^{th}$ asset and $cov(i,j)$ is the sample covariance between the return series for firms $i$ and $j$. 

We solve the mean-variance problem 
using an optimisation library that uses quadratic programming \citep{Martin2021}. Intuitively, the model is expected to have a preference for lowly correlated firms to achieve diversification benefits. A high-dimensional portfolio poses computational challenges, however, which cause our model run times to increase considerably.

\subsubsection{Network index model} \label{subsec:Network index model} \hfill\\
\indent As previously discussed studies have shown, portfolios invested in peripheral assets tend to outperform portfolios containing more central firms. One of our benchmark models, referred to as the network index model, uses the peripherality measure defined in Section \ref{subsec:Network Measures} and allocates capital according to the inverse of this score. We rescale these weights so that they sum up to one. Thus, for each node or asset $i$ in the network,
$$ w_i = 1/p_i$$
$$ w_i = \frac{w_i}{\sum w_j}.$$ 
 
The model takes a graph as input at each iteration and calculates the weights as shown above. To test this approach's performance, the resulting portfolio's daily returns are then calculated using the observed individual firm performance over the next three months. 

\subsubsection{Equal-weight portfolio} \label{subsec:Equal Weight Portfolio} \hfill\\
\indent The equal-weighted portfolio is an important benchmark strategy against which to compare any models with high dimensionality, such as those in our mid-cap universe. The strategy consists of simply assigning equal weights to all the firms in the portfolio. For each firm $i$ in a set of firms $\mathcal{V}_{t}$, the weight $w_i$ thus corresponds to

\begin{equation} \label{eq:equal}
    w_{i} = |\mathcal{V}_{t}|^{-1}.
\end{equation}

This allocation may not be practical when developing a portfolio strategy for a large number of firms, as the transaction costs increase considerably. For simplicity, however, we assume that there are no transaction costs\footnote{As discussed earlier, our method would perform well in the presence of transaction costs, since it is designed to produce a sparser portfolio. Later results will show that this no-cost assumption actually favours our benchmarks as their portfolio turnover is higher.}, and use this equal-weight portfolio as the market benchmark in our study. 
It is widely reported in the literature that the equal-weighted strategy, is difficult to beat, especially as the portfolio size increases, because the risk of model misspecification error increases for models that use complicated strategies \citep{DeMiguel2009a}.

\section{Experimental setup} \label{sec:ExperimentalSettings2}

In this section, we elaborate on how we set up our analyses and compared the different models. 

\subsection{Empirical analysis: overview and training settings} \label{subsec:overallarchitecture}
\begin{figure}[htbp]
\includegraphics[width=\textwidth]{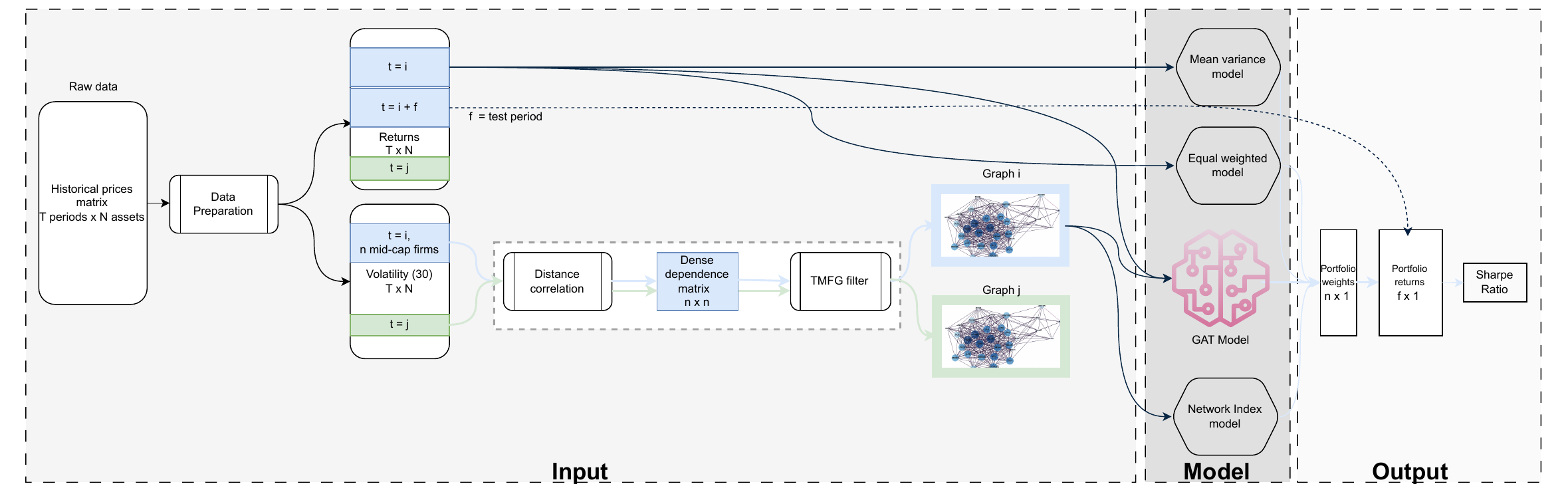}
\centering
  \caption{Empirical setup: overview}
   \label{fig:Experiment setup}
\end{figure}

At each step of our rolling window procedure, we go through a series of steps preparing the data and building and comparing our different models. This pipeline is depicted in Figure \ref{fig:Experiment setup}.  All models are measured on the same test period, and using the same set of firms. 
The equal weighted portfolio, for example, does not need training, but to facilitate a meaningful comparison, this strategy is assigned the same test set as the GAT-based model.

The first stage is the preparation of model inputs, which sees the raw data being converted to returns and volatilities. For each time period, the volatility data is then used to create the dense distance correlation matrix, which is filtered by the TMFG algorithm to create a sparse graph. 

The second stage in the pipeline is when the models receive the appropriate inputs for training. The mean-variance model and equal-weighted model are fed the expected returns data. The GAT model, in addition to the return series, also receives a filtered graph as additional input. When training this model, we adopt early stopping to prevent overfitting. A patience of 15 epochs is applied to avoid local minima. Lastly, the network index model is given the graph input, and calculates the peripherality score for each node. 

All models generate investment weights at the final output stage, and their performance is measured using (unseen) test data, which is two quarters ahead of training data and one quarter ahead of the validation data (i.e., all models are tested on an out-of-time sample). The Sharpe ratio is calculated at portfolio level for each quarter. These steps are repeated as we slide to the next window of returns, resulting in an updated set of correlation values and a new graph, which are then fed into another model run. 
We track the performance of each of these series of models over time and report the results in Section \ref{sec:Results2}.

\subsection{Evaluation metrics} \label{sec:Evaluation metrics}
The Sharpe ratio provides a suitable measure of portfolio performance. However, to further explore how the respective portfolio solutions differ, we employ some additional metrics. 

Firstly, we use two of the node centrality measures discussed earlier, specifically betweenness and degree centrality, to calculate portfolio-level centrality scores. For each of these scores, we take the weighted average over all firms in the portfolio, using the allocation percentage as respective weights. This will show how peripheral the nodes selected by each model tend to be. 

Secondly, we compare the industry-level composition of the portfolios, as different strategies might overweight certain industry sectors.We calculate each sector weighting
by summing the portfolio weights of each holding in that sector. 

Lastly, bearing in mind the dynamic nature of the problem, we also report turnover statistics. In theory, there are four possibilities for each position in a portfolio: either it is newly added, unchanged, closed, or modified (i.e.\@ its allocation increased or decreased), compared to the previous period. 
Due to the size of the mid-cap universe and its natural turnover, we chose to focus on two types of changes. Specifically, we will consider the newly added or closed positions in a portfolio, relative to the natural rate of change. The latter can be easily derived from the equal-weight portfolio, as this will create or close positions only if the companies are new to or have exited the mid-cap universe, respectively. Thus, for each model, we calculate the number of new (closed) positions in the portfolio, subtract from this the number of new (close) positions found in the equal-weight portfolio, and divide by portfolio size. We then define relative portfolio turnover as the sum of both percentages. 
 
\section{Results and discussion} \label{sec:Results2}
We begin this section by reporting the mean performance of each model according to the Sharpe ratio, followed by how this ratio varied over a 30-year period. Subsequently, we explore how the resulting portfolios differed in terms of the peripherality of the chosen holdings, sector distribution, and turnover.

\subsection{Model performance}
Table~\ref{tab:Sharpetable} shows the mean portfolio performance of the four strategies, over the training, validation and test data splits.  We annualised the shape ratio as we have a horizon of one quarter. On the test data, the GAT models have the highest Sharpe ratio of 1.082. In other words, the GAT models tend to offer better risk-adjusted returns than all of our benchmark methods. This suggests that deep learning models can learn intricate relationships based on the provided input features and the adjacency information derived from the volatility networks, which allows them to beat our benchmark strategies. In line with earlier studies confirming that even well-designed portfolio models find it hard to outperform equal-weighted portfolios (especially when the portfolios are large) \citep{DeMiguel2009a}, we can see the latter having the second-highest Sharpe Ratio, closely followed by the network index benchmark. The Mean-Variance model exhibits the worst performance, confirming similar findings in studies on large-scale portfolio optimisation \citep{Ao2019}. One explanation may be that their inputs, i.e.\@ the expected returns and (linear) covariance matrix,  
do not have the same amount of information contained in them as do the returns time series and filtered adjacency matrix that serve as the inputs to the GAT models.

\begin{table}[htbp]
  \centering
  \caption{Portfolio optimisation performance results}
  \begin{tabular}{@{}lccccl@{}}\toprule
    Sharpe Ratio (annualised) & train & val   & test \\
    \midrule
    Equal & 0.825 & 0.925 & 0.830 \\
    \midrule
    Network & 0.817 & 0.917 & 0.820 \\
    \midrule
    Mean-Variance Portfolio & 0.779 & 0.785 & 0.700 \\
    \midrule
    \textbf{GAT} & \textbf{1.819} & \textbf{1.480} & \textbf{1.082} \\
    \bottomrule
    \end{tabular}%
  \label{tab:Sharpetable}%
\end{table}%

\begin{figure}[htbp]
\includegraphics[width=\textwidth,scale=1.58]{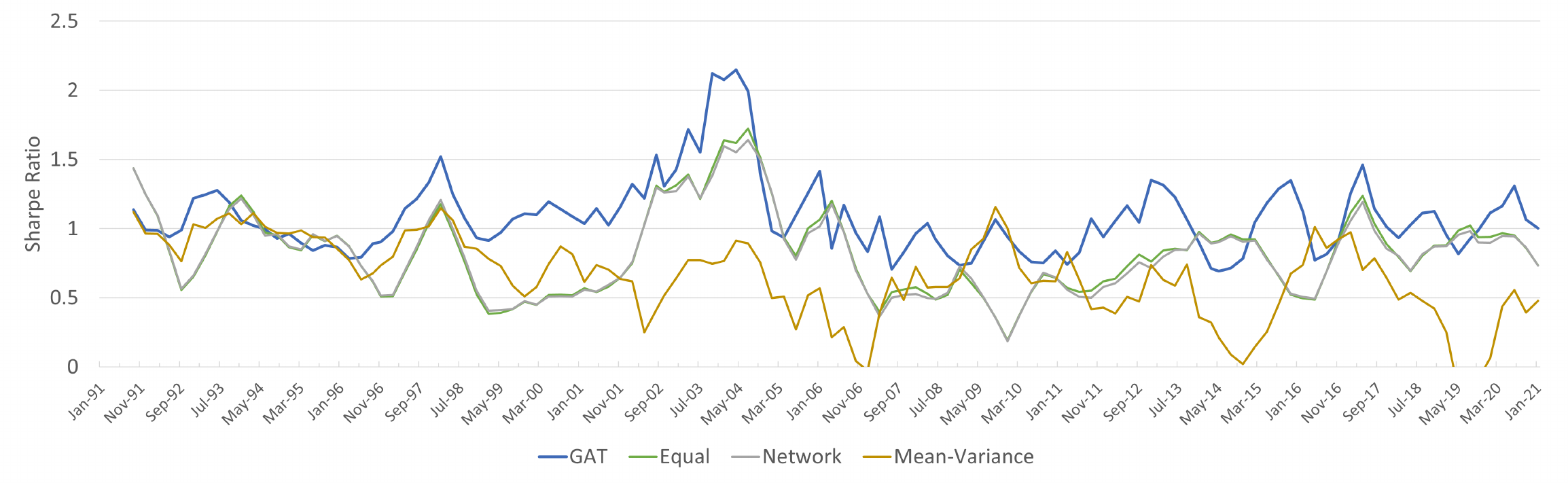}
\centering
  \caption{Model performance over time}
   \label{fig: SharpeOverTime}
\end{figure}

Whereas the rightmost column of Table~\ref{tab:Sharpetable} reports the average ratio over all the test data sets, further insights can be gained from plotting a four-quarter moving average of these quarterly ratios (showing the last 12 months' performance). 
Figure~\ref{fig: SharpeOverTime} details how the models thus perform over time. 
Apart from an initial period of five years over which all the models appear to perform similarly, this plot shows the GAT models performing consistently well from there on. We can also observe that the mean-variance models, including in the recent past, have underperformed. 
Although the relative performance gap between the GAT models and the equal-weight strategy
and network-index based models appears smaller (especially so near the end of the time span), 
the GAT models tend to more often have the edge over these two strategies as well. 
As the market went through several cycles during this extended time period, these findings appear robust to general market conditions and regime changes. 

\subsection{Strategy differences: peripherality of holdings} \label{subsec: Model differences}

\begin{figure}[hbtp]
\includegraphics[width=0.6\textwidth]{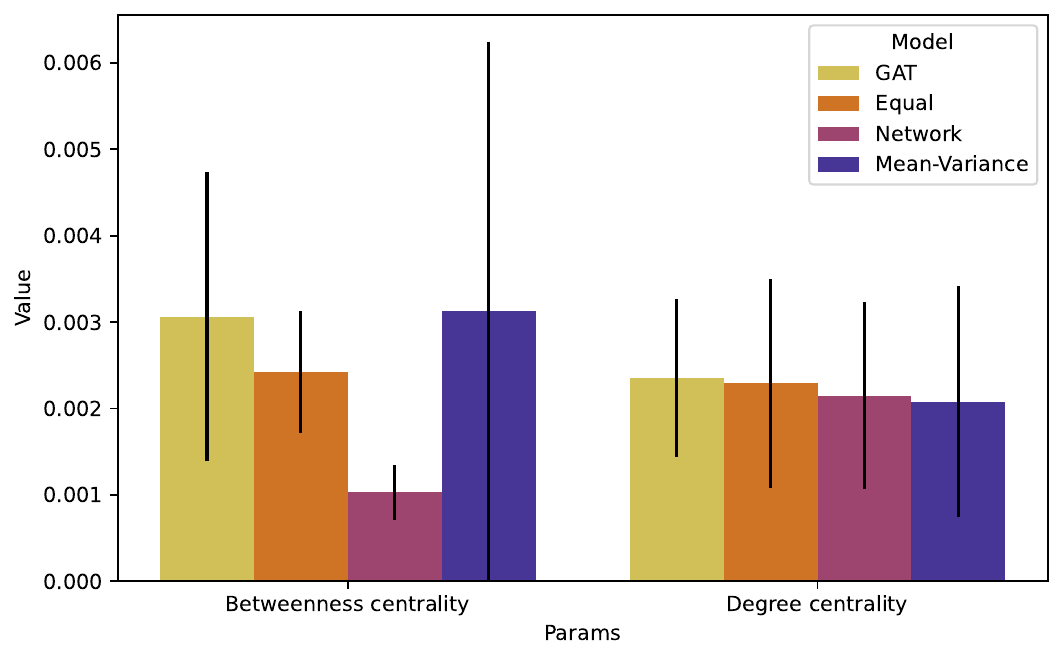}
\centering
  \caption{Weighted centrality of portfolios}
   \label{fig: ModelNetworkScore}
\end{figure}

Next, we further examine some of the factors that may explain the performance differences observed above. Firstly, Figure~\ref{fig: ModelNetworkScore} shows how two of the (weighted) average centrality scores defined earlier differ between the portfolios produced by each strategy. The set of bars on the left (right) shows betweenness (degree) centrality, respectively. The black lines represent the standard deviation of these scores over time. 

Although the GAT and mean-variance models appear similar in terms of their mean betweenness centrality, this score varies heavily over time for the mean-variance model, showing that the latter strategy is undecided in choosing between nodes that are either more or less peripheral. The betweenness scores for the GAT model show much lower variability over time, suggesting that the latter more consistently prefers companies from certain parts of the graph structure.  
The network index benchmark model chooses the most peripheral nodes according to the same criterion, but the previous section showed that the resulting portfolios underperform against those selected by the GAT model. This could be due, in part, to a relative overweight on peripheral nodes. Also, the earlier chart in Figure~\ref{fig: Default evolution} showed that mid-cap firms carry around 1.5\% default risk, which will impact the portfolio quite drastically and might eliminate any possible performance gains from selecting the more peripheral nodes. This is a different conclusion from previous studies, especially those based on topological information for portfolio optimisation \citep{Li2019}. However, those studies often excluded firms that defaulted over the study period and also tended to remove any firms that do not have the complete data available. 

From the degree centrality scores shown on the right, we can draw a similar conclusion; i.e.\@, the GAT model is associated with the smallest variance in centrality. Although the actual means are fairly close, the GAT model tends to select nodes that have marginally higher degree centrality than those in the equal-weighted portfolios or those for the other benchmarks. 

\begin{figure}[ht]
\includegraphics[width=\linewidth]{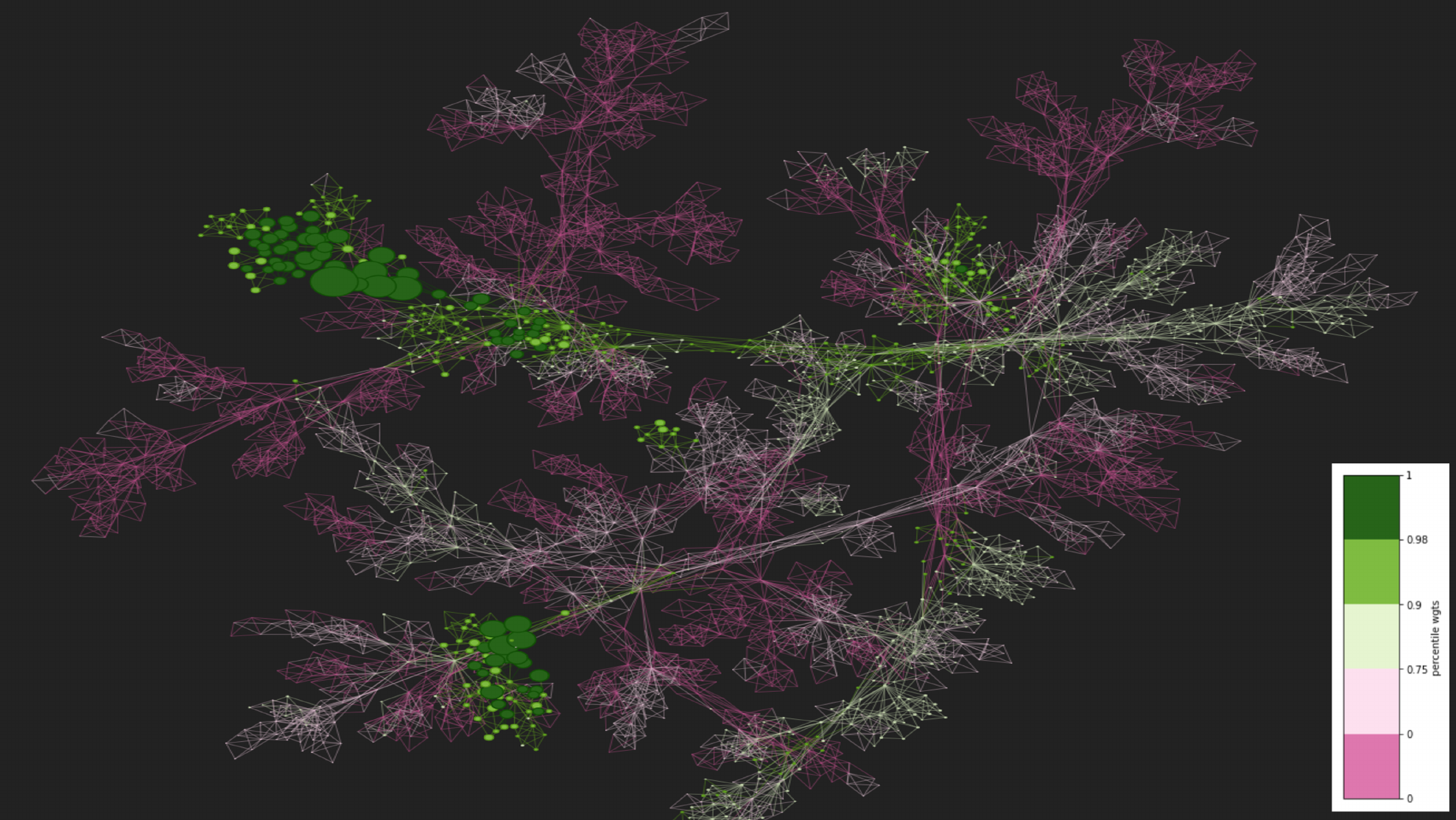}
\centering
  \caption{Distribution of GAT portfolio weights over nodes of mid-cap network snapshot}
   \label{fig: Full Graph with allocation}
\end{figure}

To shed further light on this, we revisit the filtered network previously shown in Figure~\ref{fig: Full Graph}, adding the portfolio weights allocated by the GAT model to Figure~\ref{fig: Full Graph with allocation}. This shows the allocations focusing on select branches instead of merely selecting the most peripheral nodes. The darker pink tones show where the model did not allocate any capital at all. As we move higher up in the colour scale, we see larger weights directed towards a few peripheral firms (see dark green tones) and smaller weights distributed across numerous central firms (white and light green tones).

\subsection{Strategy differences: sector allocation}

\begin{figure}[h]
\includegraphics[width=\textwidth]{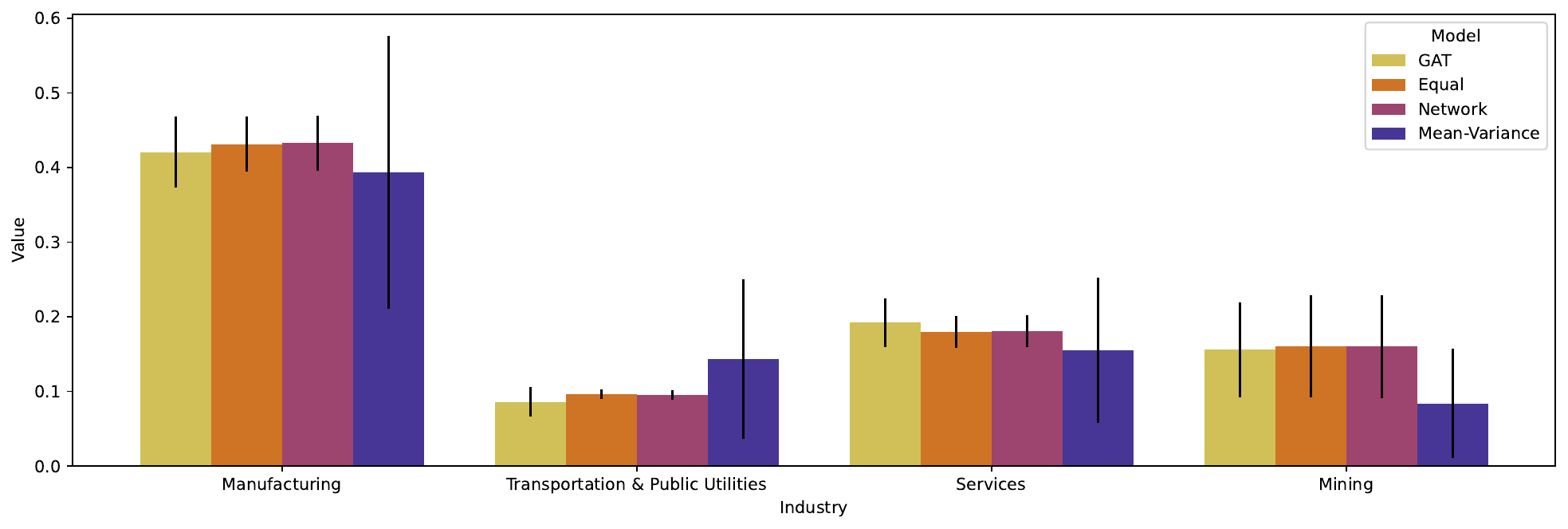}
\centering
  \caption{Industry allocations per model}
   \label{fig: ModelIndustry}
\end{figure}

Figure~\ref{fig: ModelIndustry} shows the share of the portfolio's capital that is allocated to each of the different industry sectors. 
The equal-weighted portfolio (orange bars) provides a useful baseline for comparison, as it shows how the overall population of mid-cap firms is distributed. The largest sector is manufacturing, with around 45\% of firms. The network-index model portfolios have a similar sector composition to the equal-weighted portfolio (but as seen in Figure \ref{fig: ModelNetworkScore}, they choose more peripheral nodes). The mean-variance model shows greater variance in industry weightings than other models. This is somewhat expected for a large-scale portfolio, as, with the universe of firms changing every quarter, mean-variance portfolio optimisation can be unstable. Here, it tends to underweight the larger manufacturing sector and overweight the transportation and public utilities sectors.
As seen from their relative variation in industry allocations, the GAT model, over time, rebalances sector weightings more extensively than the equal and network-based portfolios, but less so than the mean-variance model. This may put the GAT model in a better position to achieve stable returns under different market conditions. 

\subsection{Strategy differences: portfolio size and turnover}

To better understand some of their cost implications, we end the analysis by examining the resulting portfolio size and turnover for each strategy. 

\begin{figure}[htbp]
\includegraphics[width=\textwidth]{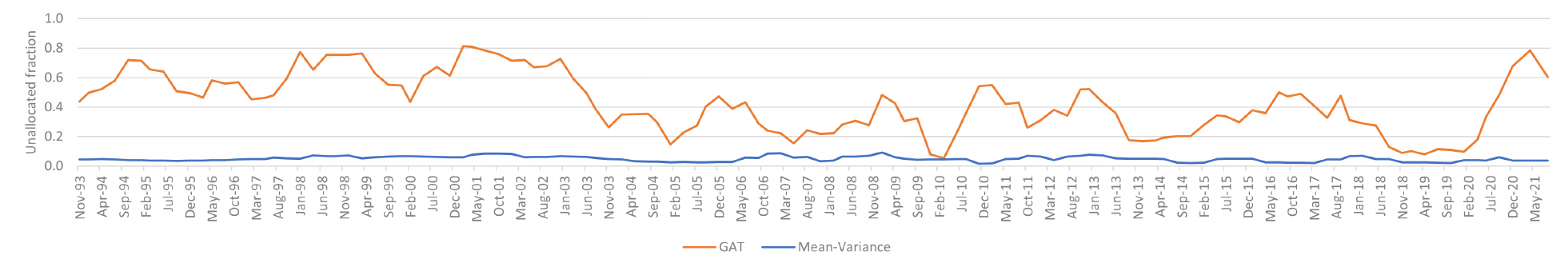}
\centering
  \caption{Fraction of firms that do not have any capital allocated, over time}
   \label{fig: allocations}
\end{figure}

Firstly, to help us compare how sparse the portfolio selections are for the GAT and mean-variance models, Figure~\ref{fig: allocations} plots, for each time period, the percentage of firms from the corresponding universe to which no capital has been allocated. Note that the network index and equal-weight strategies were left out from this chart, as both will assign non-zero weights to all of the firms. As can be clearly seen, the GAT model requires holding far fewer firms than our mean-variance implementation, which is unsurprising given that the former includes a regularisation mechanism that is lacking in the latter. The resulting difference in unallocated firms is substantial and consistent over time. The changes over time observed for the GAT model may be indicative of changing market conditions.  

\begin{figure}[tbp]
\includegraphics[width=0.7\textwidth]{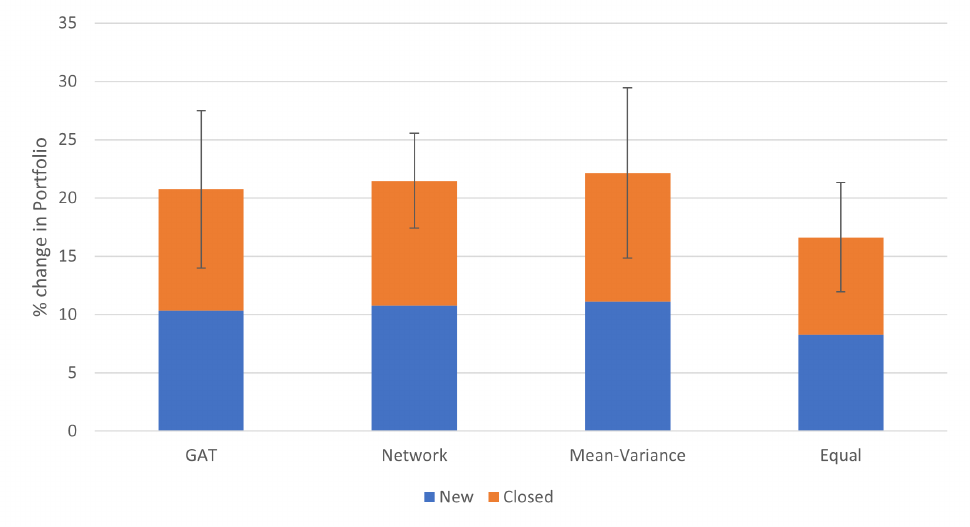}
\centering
  \caption{ Trades in the portfolio as a \% of total firms}
   \label{fig: Turnover}
\end{figure}

Secondly, we evaluate the quarterly turnover of each strategy. Unlike the capital allocation reported earlier, this relates to the count of positions changing between periods (or number of trades). Figure~\ref{fig: Turnover} shows, for the different types of portfolios, the mean proportion of holdings that were new positions, as well as how many positions were closed since the last quarter as a percentage of portfolio size. Turnover is the sum of these two, represented by the combined bar height. The equal-weight portfolio sees 16\% turnover on average, with some further variability indicated by the error bars. This reflects the natural churn as companies enter or leave the universe. 

Among the other methods, the GAT model has the lowest turnover, albeit with a higher variance than the network model. This means that, on average, it requires fewer trades but, depending on market conditions, there are some periods where it uses substantially more than the network index model. 

Table~\ref{tab:Turnover} summarises how much of each model's turnover is in excess of the natural change (by subtracting the turnover of the equal-weight portfolio). 
\begin{table}[b]
  \centering
  \caption{Mean (excess) turnover by model (in excess of natural change)}
    \begin{tabular}{@{}lccccl@{}}
    \toprule
    \% change & new & closed & turnover \\
    \midrule
    Network  & 2.48  & 2.36  & 4.84 \\
    \midrule
    Mean-Variance Portfolio    & 2.83  & 2.68  & 5.50 \\
    \midrule
    \textbf{GAT} & \textbf{2.07}  & \textbf{2.03} & \textbf{4.10} \\
    \bottomrule 
    \end{tabular}%
  \label{tab:Turnover}%
\end{table}%
From this, we can again see that the GAT model requires little such excess turnover compared to the other strategies. For example, the portfolios produced by the GAT model, on average, add a mere 2.07\% more new positions from one quarter to the next, whilst they close an additional 2.03\% of existing positions (compared to 2.83\% and 2.68\% for the mean-variance portfolios). With an overall excess turnover of 4.10\%, the GAT model is thus associated with a relative reduction of 25\% and 15\%, compared to the mean-variance and network index models, respectively.
This strongly suggests that the GAT model will have lower transaction costs compared to traditional models.

\section{Conclusion} \label{sec:Conclusion2}

In this paper, we have put forward a solution for large-scale portfolio optimisation using deep graph learning.
We have seen how GAT-based models, a type of model within the broader family of deep learning models, can extract intricate relationships that other traditional models can not. 
While most studies focus on portfolio optimisation for assets that have regular availability of prices, we focused on problems where the data is more difficult to model, by choosing to analyse the volatile mid-cap market. The study of such firms in itself is of interest for a variety of reasons. Whilst they are far more numerous than their large-cap counterparts, their smaller size makes them more vulnerable to market movements and larger correlation with the overall economy, something most structured stochastic models ignore. They also potentially offer higher returns for commensurate risk to investors, while better risk management may lead to better access to financial markets for those firms. 
      
In designing our approach, we have linked several areas of study. We applied the distance correlation measure to firm volatility pairings, to capture more complex connections between firms than with alternative approaches. From this, we generated a sparse graph by employing the Triangulated Maximally Filtered Graph algorithm, a filtering technique that is applicable to large-scale graphs. Through this, we explicitly incorporate the interdependence of midcap companies. These filtered graphs were then presented to a GAT model, which can identify higher-order relationships. The final allocation layers of our deep learning solution were designed to optimise the embeddings generated by the GAT models, and the regularisation parameters used in the deep learning models imposed constraints on possible weights and the number of firms to which capital can be allocated. Being derived from the Sharpe ratio, the chosen loss function set a risk-adjusted return objective for portfolio performance maximisation. Other portfolio objectives could similarly be used and further constraints imposed on the portfolio allocations.  

Starting from the premise that deep learning models should be adept at optimising many high-dimensional problems such as ours, our experiments with real-world midcap data indeed showed that the GAT-based models achieved better performance than other alternatives. We also studied how these results were robust to different market conditions and looked at the distribution of firm allocations across different strategies, to identify some of the factors explaining how the GAT models differed.
In so doing, we found that they tended to choose companies that are not too much in the periphery and allocated capital to fewer firms. Lastly, we observed that the typical turnover associated with the GAT models was lower than that of the alternatives, although, on fewer occasions, they did make more substantial changes to reposition the portfolio. 
 
As for future work, further graph neural network models could be developed to predict aspects like market regimes, or produce early-warning indicators for financial networks. By changing the objectives and revising the loss function, we could also extend deep graph learning-based portfolio optimisation models towards different goals, such as the construction of Environmental, Social and Governance (ESG) portfolios, or other types of diversified portfolios, and solve problem instances on a much larger scale than before.

\section*{Acknowledgements}
This work was supported by the Economic and Social Research Council [grant number ES/P000673/1]. The last author acknowledges the support of the Natural Sciences and Engineering Research Council of Canada (NSERC) [Discovery Grant RGPIN-2020-07114]. This research was undertaken, in part, thanks to funding from the Canada Research Chairs program.

The authors acknowledge the use of the IRIDIS High Performance Computing Facility, and associated support services at the University of Southampton, in the completion of this work.

\begin{appendices}

\end{appendices}

\end{document}